\documentclass[fleqn,usenatbib]{mnras}
\usepackage{graphicx}	
\usepackage{amsmath}	
\usepackage{amssymb}	
\usepackage{multicol}	
\usepackage{bm}		

\newcommand{\redmapper}{\textit{redMaPPer}}
\newcommand{\redmagic}{\textit{redMaGiC}}
\newcommand{\sdss}{\textit{SDSS}}
\newcommand{\des}{\textit{DES}}

\newcommand{\sv}{\des\ \textit{SV}}

\newcommand{\ngmix}{\texttt{NGMIX}}
\newcommand{\photoz}{photo-$z$}

\newcommand{\PHOTOZS}{PHOTO-$z$'s}
\newcommand{\photozs}{photo-$z$'s}
\newcommand{\yearfive}{\des\ Y5}

\newcommand{\ccz}{cross-correlation}

\newcommand{\cczs}{cross-correlations}
\newcommand{\Cczs}{Cross-correlations}

\newcommand{\Ru}{\mathrm{R_{u}}}
\newcommand{\Rr}{\mathrm{R_{r}}}
\newcommand{\Du}{\mathrm{D_{u}}}
\newcommand{\Dr}{\mathrm{D_{r}}}
\newcommand{\DD}{\Du \Dr}
\newcommand{\RR}{\Ru \Rr}
\newcommand{\DRu}{\Du \Rr}
\newcommand{\DRr}{\Ru \Dr}
\newcommand{\NDD}{N_{\Du} N_{\Dr}}
\newcommand{\NRR}{N_{\Ru} N_{\Rr}}

\newcommand{\be}{\begin{equation}}
\newcommand{\ee}{\end{equation}}

\newcommand{\bea}{\begin{eqnarray}}
\newcommand{\eea}{\end{eqnarray}}

\newcommand{\Hinv}{H^{-1}}

\newcommand{\avg}[1]{\langle #1 \rangle}

\usepackage[usenames]{color}
\definecolor{purple}{RGB}{150,0,200}

\newcommand{\lowz}{{\it LOWZ}}
\newcommand{\cmass}{{\it CMASS}}

\newcommand{\Delu}{\Delta_{\rm{u}}}
\newcommand{\Delref}{\Delta_{\rm{ref}}}

\newcommand{\pu}{\phi_{\rm{u}}}
\newcommand{\pref}{\phi_{\rm{ref}}}

\newcommand{\bu}{b_{\rm{u}}}
\newcommand{\bref}{b_{\rm{ref}}}

\newcommand{\zref}{z_{\rm{ref}}}

\newcommand{\wm}{w_{\rm{mm}}}

\newcommand{\Rmin}{r_{\rm{min}}}
\newcommand{\Rmax}{r_{\rm{max}}}

\newcommand{\Mpc}{\rm{Mpc}}

\usepackage[T1]{fontenc}
\usepackage{ae,aecompl}

\usepackage{txfonts}

\makeatletter
\def\blfootnote{\xdef\@thefnmark{}\@footnotetext}
\makeatother

\title[DESSV Photo-$z$ Calibration with Clustering-$z$]{Cross-Correlation Redshift Calibration Without Spectroscopic Calibration Samples in DES Science Verification Data}

\usepackage{eso-pic}

\AddToShipoutPictureBG*{%
  \AtPageUpperLeft{%
    \hspace{0.75\paperwidth}%
    \raisebox{-3.5\baselineskip}{%
      \makebox[0pt][l]{\textnormal{DES 2016-0206}}
}}}%

\AddToShipoutPictureBG*{%
  \AtPageUpperLeft{%
    \hspace{0.75\paperwidth}%
    \raisebox{-4.5\baselineskip}{%
      \makebox[0pt][l]{\textnormal{FERMILAB-PUB-17-284-AE}}
}}}%

\AddToShipoutPictureBG*{%
  \AtPageUpperLeft{%
    \hspace{0.75\paperwidth}%
    \raisebox{-5.5\baselineskip}{%
      \makebox[0pt][l]{\textnormal{SLAC-PUB-17113}}
}}}%

\author[DES Collaboration]{
\parbox{\textwidth}{
\Large
C.~Davis$^{1}$\thanks{E-mail: cpd@stanford.edu (CPD) },
E.~Rozo$^{8}$,
A.~Roodman$^{1,7}$,
A.~Alarcon$^{2}$,
R.~Cawthon$^{3}$,
M.~Gatti$^{4}$,
H.~Lin$^{5}$,
R.~Miquel$^{6,4}$,
E.~S.~Rykoff$^{1,7}$,
M.~A.~Troxel$^{9,10}$,
P.~Vielzeuf$^{4}$,
T. M. C.~Abbott$^{11}$,
F.~B.~Abdalla$^{12,13}$,
S.~Allam$^{5}$,
J.~Annis$^{5}$,
K.~Bechtol$^{14}$,
A.~Benoit-L{\'e}vy$^{15,12,16}$,
E.~Bertin$^{15,16}$,
D.~Brooks$^{12}$,
E.~Buckley-Geer$^{5}$,
D.~L.~Burke$^{1,7}$,
A. Carnero Rosell$^{17,18}$,
M.~Carrasco~Kind$^{19,20}$,
J.~Carretero$^{4}$,
F.~J.~Castander$^{2}$,
M.~Crocce$^{2}$,
C.~E.~Cunha$^{1}$,
C.~B.~D'Andrea$^{21}$,
L.~N.~da Costa$^{17,18}$,
S.~Desai$^{22}$,
H.~T.~Diehl$^{5}$,
P.~Doel$^{12}$,
A.~Drlica-Wagner$^{5}$,
A.~Fausti Neto$^{17}$,
B.~Flaugher$^{5}$,
P.~Fosalba$^{2}$,
J.~Frieman$^{5,3}$,
J.~Garc\'ia-Bellido$^{23}$,
E.~Gaztanaga$^{2}$,
D.~W.~Gerdes$^{24,25}$,
T.~Giannantonio$^{26,27,28}$,
D.~Gruen$^{1,7}$,
R.~A.~Gruendl$^{19,20}$,
G.~Gutierrez$^{5}$,
K.~Honscheid$^{9,10}$,
B.~Jain$^{21}$,
D.~J.~James$^{29,11}$,
T.~Jeltema$^{30}$,
E.~Krause$^{1}$,
K.~Kuehn$^{31}$,
S.~Kuhlmann$^{32}$,
N.~Kuropatkin$^{5}$,
O.~Lahav$^{12}$,
T.~S.~Li$^{5}$,
M.~Lima$^{33,17}$,
M.~March$^{21}$,
J.~L.~Marshall$^{34}$,
P.~Martini$^{9,35}$,
P.~Melchior$^{36}$,
R.~L.~C.~Ogando$^{17,18}$,
A.~A.~Plazas$^{37}$,
A.~K.~Romer$^{38}$,
E.~Sanchez$^{39}$,
V.~Scarpine$^{5}$,
R.~Schindler$^{7}$,
M.~Schubnell$^{25}$,
I.~Sevilla-Noarbe$^{39}$,
M.~Smith$^{40}$,
M.~Soares-Santos$^{5}$,
F.~Sobreira$^{41,17}$,
E.~Suchyta$^{42}$,
M.~E.~C.~Swanson$^{20}$,
G.~Tarle$^{25}$,
D.~Thomas$^{43}$,
V.~Vikram$^{32}$,
A.~R.~Walker$^{11}$,
R.~H.~Wechsler$^{44,1,7}$
\begin{center} (DES Collaboration) \end{center}
}
}
\vspace{0.4cm}

\begin{document}
\label{firstpage}
\pagerange{\pageref{firstpage}--\pageref{lastpage}}
\maketitle

\begin{abstract}
  Galaxy cross-correlations with high-fidelity redshift samples hold the potential to precisely calibrate systematic photometric redshift uncertainties arising from the unavailability of complete and representative training and validation samples of galaxies.  However, application of this technique in the Dark Energy Survey (\des) is hampered by the relatively low number density, small area, and modest redshift overlap between photometric and spectroscopic samples.  We propose instead using photometric catalogs with reliable photometric redshifts for \photoz\ calibration via \cczs.  We verify the viability of our proposal using \redmapper\ clusters from the Sloan Digital Sky Survey (\sdss) to successfully recover the redshift distribution of \sdss\ spectroscopic galaxies.  We demonstrate how to combine \photoz\ with \ccz\ data to calibrate photometric redshift biases while marginalizing over possible clustering bias evolution in either the calibration or unknown photometric samples.  We apply our method to \des\ Science Verification (\sv) data in order to constrain the photometric redshift distribution of a galaxy sample selected for weak lensing studies, constraining the mean of the tomographic redshift distributions to a statistical uncertainty of $\Delta z \sim \pm 0.01$.  We forecast that our proposal can in principle control photometric redshift uncertainties in \des\ weak lensing experiments at a level near the intrinsic statistical noise of the experiment over the range of redshifts where \redmapper\ clusters are available. Our results provide strong motivation to launch a program to fully characterize the systematic errors from bias evolution and \photoz\ shapes in our calibration procedure.
\end{abstract}
\blfootnote{Affiliations are listed at the end of the paper.}
\setcounter{footnote}{1}


\section{Introduction}
\label{sec:intro}

The determination of the photometric redshift distribution of a given source population is
fraught with difficulties.
For instance, the two primary methods for determining the redshift distribution of photometric objects --- template fitting and machine learning --- must both confront a critical difficulty: the spectroscopy available for training, calibration, and validation of photometric redshift
techniques is rarely representative of the
magnitude and color-space distribution of all survey galaxies.
It is possible to mitigate this problem by weighting spectroscopic galaxies such that they better represent the photometric properties of the whole photometric survey \citep[][Hoyle et al. in prep.]{Lima:2008aa, Cunha:2009aa,Sanchez:2014aa,Bonnett:2015aa}. 
However, for this method to be effective, the training set must be complete relative to the photometric data, such that it densely covers and spans the same space of relevant photometric observables as the full survey. This is difficult to achieve in regions that lack spectroscopic data, particularly at high redshifts.
Similarly, available templates may not span the full color-redshift-space
\citep{Bonnett:2015aa} of the galaxies of interest. This problem tends to be particularly acute for the faint galaxies which make up the majority of the sources used for weak gravitational lensing.
In order to precisely and accurately determine
the dark energy equation of state from photometric weak
lensing and large-scale structure measurements, it is vital to
precisely characterize the redshift distributions of
the tomographic redshift bins into which the galaxies are split.
\citet{Bonnett:2015aa} find that photometric
redshifts biases must be controlled at the $\sim \! 0.003$ level in order for
the Dark Energy Survey (\des) 5,000 $\deg^2$ survey to not be limited by photometric redshift uncertainties.
While `\photoz' methods
have made considerable progress towards meeting these requirements,
current performance falls short of this goal.
As such, any method with the potential to further improve this calibration
is of great interest as a possible way to reduce, and perhaps even eliminate,
the \photoz\ systematics floor of photometric surveys like the \des.

\citet{Newman:2008aa} was the first to demonstrate that by cross-correlating a
sample of photometric galaxies with unknown redshift distribution with thin redshift
slices of a spectroscopic galaxy sample one could recover the redshift
distribution of the photometric galaxy sample.  This method was improved by
\citet{Matthews:2010aa} using an iterative technique to account for the
evolution in galaxy clustering bias.  Several others have tested the method on N-body
simulations with promising results, including various methods
for further improving and refining the cross-correlation method
\citep{Schulz:2010aa, Matthews:2012aa, Schmidt:2013ac, McQuinn:2013aa,McLeod:2016aa,Scottez:2017aa,van-Daalen:2017aa}.
The method has also been applied to data: \citet{Benjamin:2010aa} used \cczs\ to measure the degree of artificial contamination in tomographic redshift bins and applied the technique to the Canada-France-Hawaii Telescope Legacy Survey (CFHTLS), and \citet{Menard:2013aa} used clustering measurements on both linear and modestly non-linear scales to characterize the redshift distributions of SDSS, WISE, and FIRST galaxies.

More recently, \citet{Rahman:2014aa} combined clustering information with photometry to show how the \ccz\ method can recover the redshift probability distribution of an individual galaxy.  
\citet{Schmidt:2015aa} cross-correlated Planck High Frequency Instrument maps against SDSS quasars to estimate the distribution of the cosmic infrared background.  
\citet{Rahman:2015aa} mapped the relation between galaxy color and redshift by using \cczs\ instead of spectral energy distribution templates, and \citet{Scottez:2016aa} apply the same estimations to VIPERS.
\citet{Sun:2015aa} extend \cczs\ to the Fourier domain by examining how galaxy angular power spectra can determine the mean redshift to percent precision.  
\citet{Lee:2016ab} use integer linear programming to optimize cross- and autocorrelations, demonstrating that it is possible to assign individual galaxies to redshift bins via clustering signals alone.
\citet{Choi:2015aa} used \cczs\ to check the validity of using summed $p(z)$ to determine galaxy redshift distributions in both CFHTLS and the Red-sequence Cluster Lensing Survey.

Most recently, \citet{Hildebrandt:2016aa} used cross-correlation methods
to validate the photometric redshift distribution of source galaxies
in the Kilo Degree Survey (KiDS) via the open source code \texttt{The-wiZZ} described in \citet{Morrison:2016aa}. \citet{Johnson:2016aa} extended the quadratic estimator from \citet{McQuinn:2013aa} and found qualitative agreement with the results from \citet{Hildebrandt:2016aa}. Importantly, those efforts
were hampered by the low number of spectroscopic galaxies available
for \photoz\ calibration via cross-correlations across a broad redshift range.
This relative lack of spectroscopic calibration samples for \ccz\
studies is a serious obstacle to the realization of the promise that
these methods hold.

One solution to the relative paucity of spectroscopic galaxies in the footprint
of these wide-field optical photometric surveys is to use other objects which
have reasonably precise redshifts but which are far more numerous.
One example for such a class of objects are \redmapper\ galaxy clusters described in \citet{Rykoff:2014aa}, who present a red sequence cluster finder that produces
objects with nearly Gaussian estimated redshifts and scatter $\sigma_z / (1 + z) \sim 0.01$. 
The method relies
on a small set
of spectroscopic objects for training, but can then be used to find objects
over the full survey footprint to much fainter magnitudes.  Importantly, while
these objects are rare, the complete overlap between these photometrically selected
objects and the galaxies that are to be calibrated means that the cross-correlation signal
can be measured with high signal-to-noise. 
This makes them a natural
candidate for calibrating redshift distributions in photometric wide-field
surveys.  Note, however, that such cross-correlation measurements are by necessity
limited to the redshift range over which the red sequence is well calibrated. 

In this paper we examine how well the \ccz\ method performs when,
instead of spectroscopic galaxies, objects with accurate photometric redshifts
are used as the reference sample. We wish to examine the following questions:
\begin{itemize}
    \item{How well do \ccz\ methods with non-spectroscopic reference samples
        perform in comparison with spectroscopic reference samples?}
    \item{How can we properly combine \photoz\ and \ccz\ information to minimize the noise inherent to \ccz\ methods while reducing the redshift biases from standard \photoz\ methods?}
    \item{How does using non-spectroscopic reference
        samples to constrain the redshift distributions of galaxies impact cosmological
        parameter estimation?}
\end{itemize}

This paper is organized as follows.
In Section~\ref{sec:datasets} we present the datasets used in this study.
In Section~\ref{sec:methods} we review the theory behind the \ccz\ method and present our method for calculating redshift distributions from \cczs.
In Section~\ref{sec:sdss} we present the performance of \cczs\ when using \redmapper\ clusters as reference objects, and compare our results to those obtained when using \sdss\ spectroscopic galaxies as reference.  We also examine the impact of using different weighting functions and integration ranges on the accuracy and precision of the recovered redshift distributions. 
In Section~\ref{sec:des} we apply our method to the Dark Energy Survey Science Verification dataset, and examine how we can use the \ccz\ method to determine the redshift bias in photometric redshift methods.  
In Section~\ref{sec:cosmo} we forecast the performance of using the \ccz\ method to constrain cosmological parameters with the Dark Energy Survey Year Five data. 
We wrap up in Section~\ref{sec:conclusions} and discuss potential future applications of this method.

Throughout this paper we assume a WMAP9 cosmology ($\Omega_m = 0.286$) and
report distances in $h^{-1}$ Mpc. We find that the choice of cosmology has a negligible impact on our results \citep{Hinshaw:2013aa}.

\section{Datasets}
\label{sec:datasets}

Our analysis relies on four catalogs drawn from two galaxy surveys, the Sloan Digital Sky Survey (\sdss) and
the Dark Energy Survey Science Verification (\sv).  Each of these data sets and the corresponding catalogs are detailed
below.

\subsection{\sdss\ Spectroscopic Galaxy Samples}
\label{sec:datasets:sdss}

The \sdss\ spectroscopic galaxy sample consists of the \lowz\ and
\cmass\ galaxy samples from the Baryon Oscillation Spectroscopic Survey (BOSS) \citep{Alam:2015aa}. BOSS obtained data
over 9376 deg$^{2}$ divided into two regions in the
Northern and Southern Galactic Caps. The \lowz\ sample is a set of
galaxies uniformly targeted for large-scale structure studies in a
relatively low redshift range ($z < 0.43$). The \cmass\ sample is
a set of galaxies over the range ($0.43 < z < 0.7$) designed to create
an approximately volume-limited sample in stellar mass.
Both catalogs come with `randoms,' catalogs that reflect the footprints and 
selection functions of the two surveys. For our purposes
here it is sufficient to simply combine the two catalogs over their full redshift ranges, which we shall
collectively refer to as the \sdss\ catalog.

\subsection{\redmapper}
\label{sec:datasets:redmapper}

\redmapper\ is a red sequence cluster finder originally developed within the
context of the \sdss\ \citep{Rykoff:2014aa,Rozo:2015ab}. 
The red sequence is the empirical relation that early-type galaxies in rich clusters lie along a linear color-magnitude relation with small scatter \citep{Bower:1992aa,Bell:2004aa}.
\redmapper\ iteratively self-trains a model of the red sequence
based on sparse spectroscopic data, and then uses this model to identify
clusters of red sequence galaxies, and to estimate the photometric redshift
of the resulting clusters.
The
algorithm has been extensively tested and validated \citep{Rozo:2014aa}.
In addition to the \sdss\ catalog, we
use the \redmapper\ cluster catalog resulting from the application
of the \redmapper\ algorithm to the \sv\ data \citep{Rykoff:2016aa}.
While the \redmapper\ cluster catalog from \sdss\ is therefore different from the one used in \sv, in that the algorithm has been applied to a different set of galaxies, the number density and the mass-richness relations are similar, and most importantly, the performance in \photoz\ is the same.
These catalogs come with their own associated `randoms' which reflect the footprint
and the ability to select a \redmapper\ cluster of some richness $\lambda$ at a
given location. We have chosen clusters with $\lambda > 5$.
For our purposes, the most important aspect of the \redmapper\ cluster
catalog is its exquisite redshift performance: cluster \photozs\
are both accurate (\photoz\ redshift biases are at the $0.005$ level or less)
and precise (\photoz\ scatter is $\sigma_z/(1+z)\sim 0.01$)
with a $4\sigma$ outlier fraction of at most 3\%.  As such, the \redmapper\
clusters are excellent candidates for photometric reference samples
when attempting to perform \photoz\ calibration via cross-correlation.

Unfortunately, the \redmapper\ clusters do not span the full redshift range of the samples we consider in this paper. In \sdss, the redshift range of the \redmapper\ clusters used is $0.1 < z < 0.56$, while in \sv\ we take advantage of the greater depth and look at clusters in the range $0.2 < z < 0.85$.  The greater redshift reach in the \sv\ region is due to the greater depth of the photometric data. In contrast, the \sdss\ spectroscopic galaxies extend to $z < 0.7$, while the source galaxies in \sv\ can have redshifts of $z > 1$.

\subsection{\sv}

The Dark Energy Survey is an ongoing $5000\ \mathrm{deg}^2$ photometric survey in the $grizY$ bands performed with the
Dark Energy Camera~\citep[DECam,][]{Flaugher:2015aa}.  Before the beginning of the
survey, the \des\ conducted a $\sim250\,\mathrm{deg}^2$ ``Science
Verification'' (SV) survey.  The main portion of the SV footprint, used in this
paper, covers $139\ \mathrm{deg}^2$ in the range
$65<\mathrm{R.A.}<93$ and $-60<\mathrm{Decl.}<-42$. The region is dubbed South Pole Telescope East (SPTE)
because of its location and overlap with the South Pole Telescope survey
area~\citep[SPT,][]{Carlstrom:2011aa}.  SPTE was observed with 2-10 tilings in each of the $griz$ filters.  In
addition, the \des\ observes 10 Supernova fields every 5-7 days, each of which covers
a single DECam 2.2 degree-wide field-of-view.  The median depth of the SV survey
(defined as $10\sigma$ detections for extended sources) are $g=24.0$, $r=23.8$
$i=23.0$, $z=22.3$, and $Y=20.8$.

The \sv\ data were processed by the \des\ Data Management (DESDM)
infrastructure~\citep[Morganson et al, in prep, see also][]{Sevilla:2011aa,Desai:2012aa}.
We used {\tt SExtractor}~\citep{Bertin:1996aa, Bertin:2011aa} to detect photometric sources
based on a ``chi-squared'' coadd of $r$, $i$, and $z$ images
obtained with {\tt SWarp}~\citep{Bertin:2010aa}.  The catalog was then further
refined to produce the \des\ ``SVA1 Gold'' catalog.\footnote{\url{https://des.ncsa.illinois.edu/releases/sva1}}
Galaxy magnitudes and colors are computed via the
{\tt SExtractor} {\tt MAG\_AUTO} quantity.   This SVA1 Gold catalog is
the fundamental input to the construction of the \sv\ \ngmix\ galaxy
sample and the \sv\ \redmapper\ cluster catalog.

\subsection{\ngmix}
\label{sec:datasets:ngmix}

From the SVA1 Gold catalog, we examine a subsample of galaxies used for cosmic shear measurements.\footnote{The \sv\ cosmic shear analysis also used a second shape measurement pipeline, \texttt{IM3SHAPE} \citep{Zuntz:2013aa}. Since we are only interested in verifying the feasibility of the proposed cross-correlation measurement, in this work we have limited ourselves to the \ngmix\ sample because of its higher 
space density.}
The \ngmix\ pipeline \citep{Sheldon:2014aa} estimates the shapes of the galaxies in the SVA1 catalog. The subsample is then selected by cutting objects with very low surface brightnesses and small sizes, and choosing only objects with reasonable colors ($-1 < g - r < 4$ and $-1 < i - z < 4$).
\ngmix\ represents galaxies as sums of Gaussians \citep{Hogg:2013aa}. The same model shape is fitted simultaneously across multiple exposures in the \textit{riz} bands using Markov Chain Monte Carlo techniques applied to a full likelihood which forward models the galaxy by convolving with the exposure PSF. The final effective source number density of the \ngmix\ catalog is $\sim 6.1$ galaxies per square arcminute \citep{Becker:2015aa}. The shape systematics in the \ngmix\ catalog are tangential to this paper to the extent that they do not imprint a spatial correlation on the footprint. However, interested readers should look to \citet{Jarvis:2016aa} for a thorough examination of shape systematics. 

Galaxy \photozs\ for the \ngmix\ catalog are estimated using four different
photometric redshift algorithms: the template-based algorithm \texttt{BPZ} \citep{Benitez:1999aa, Coe:2006aa},
and the machine learning algorithms \texttt{SkyNet} \citep{Graff:2014aa, Bonnett:2015ab},
\texttt{TPZ} \citep{Carrasco-Kind:2013aa, Carrasco-Kind:2014ad}, and \texttt{ANNZ2} \citep{Sadeh:2015aa}.
Extensive testing of the algorithms was carried out in \citet{Sanchez:2014aa}
and \citet{Bonnett:2015aa}, to which we refer the reader for further detail
on the algorithms. We do not include the calibration offset to \texttt{BPZ} that was measured in \citet{Bonnett:2015aa}. For our purposes, the key point is that each of these algorithms
return a redshift probability distribution $P(z)$ for each galaxy, the sum of which
can be used as a (biased) estimator for the redshift distribution of
the galaxies under consideration \citep{Asorey:2016aa}.


\section{Theory and Methods}
\label{sec:methods}

We outline the theory in this section and present our estimator. The
approach presented here is based on \citet{Menard:2013aa}.
We repeat much of the argument here, though the focus and presentation are
somewhat different.  Our goal is not to be repetitious, but to simply present
an alternative but fully equivalent discussion.

Let $\rho$ be the comoving galaxy density.  In a flat universe, the corresponding galaxy density per unit angular area per unit redshift
$d^2 N/d\Omega dz$ is given by
\be
\frac{d^2N}{d\Omega dz} (d\Omega dz) =  \rho(z) \chi^2(z) \frac{d\chi}{dz} d\Omega dz = \rho(z) \chi^2(z) c\Hinv(z) d\Omega dz .
\ee
where $\chi$ is the comoving distance and $H$ is the Hubble parameter. One has then
\be
\nu \equiv \frac{d^2 N}{d\Omega dz}  =  \bar \rho(z) \chi^2(z) c\Hinv(z)  \left[ 1 + \delta(\chi(z)\theta,\chi(z)) \right] \\
\ee
where $\nu$ is the angular galaxy density per unit redshift and per unit angular area, $\delta$ is the
fractional matter overdensity relative to the mean density, and $\theta$ is the angular separation.

Let $n(\theta) = dN/d\Omega$ be the projected galaxy density. One has
\be
n(\theta) = \int dz\ \bar \rho(z) \chi^2(z) c\Hinv(z)  \left[ 1 + \delta(\chi(z)\theta,\chi(z)) \right].
\ee
The mean galaxy density is simply
\be
\bar n = \int dz\ \bar \rho(z) \chi^2(z) c\Hinv(z),
\ee
while the redshift distribution of the galaxies $\phi(z)$ is defined such that $\bar n \phi(z)$ is the
mean galaxy density per unit redshift, i.e.
\be
\bar n \phi(z) = \bar \nu = \bar \rho(z) \chi^2(z) c\Hinv(z)
\ee
We can therefore write
\be
n(\theta) = \bar n \int dz\ \phi(z) \left[ 1 + \delta(\chi(z)\theta,\chi(z)) \right]
\ee
where $\phi(z)$ integrates to unity.  Setting $n(\theta) = \bar n (1+\Delta)$, we see that
\be
\Delta (\theta) = \int dz\ \phi(z) \delta(\chi\theta,\chi).
\ee

Consider now two galaxy samples, one of which has known redshifts, which we refer to as
the \it reference \rm sample with subscript `ref,' and one which has an unknown redshift distribution, which
we refer to as the \it unknown \rm sample with subscript `u.'  We wish to consider the cross-correlation
between the unknown sample and reference galaxies within a narrow redshift bin.
The angular cross-correlation between the reference and unknown samples is therefore
\begin{align}
w(\theta) &= \avg{ \Delu\Delref }(\theta) \\ &= \int dz dz'\ \pu(z)\pref(z') \bu(\theta, z)\bref(\theta, z') \xi( \chi\theta,\chi; \chi'\theta,\chi' )
\end{align}
where $\xi$ is the matter--matter correlation function $\avg{\delta(\chi\theta,\chi) \delta(\chi'\theta,\chi')}$ and we allow the galaxy clustering bias $b$ to have both redshift and scale dependence for some separation $\theta$ and redshift $z$.
Consider the case $\pref(z) = \delta(z-\zref)$ where $\zref$ is some reference redshift. This is equivalent to selecting a reference sample in an infinitely narrow redshift slice. We have then
\be
w(\theta, \zref) = \bref(\theta, \zref) \int dz\ \pu(z) \bu(\theta, z) \xi( \chi\theta,\chi; \chi'\theta,\chi' ) .
\ee
Now, $\xi$ is zero unless $\chi\sim \chi'=\chi(\zref)$.  Using a flat sky approximation, and adopting
the origin at a reference galaxy when measuring the angular separation we find
\begin{align}
w(\theta, \zref) = \bref(\theta, \zref) \int dz\ & \pu(z) \bu(\theta, z) \ \times \nonumber \\ &\xi \left( \left[ \Delta \chi^2(z, \zref) + \chi(\zref)^2 \theta^2 \right]^{1/2} ; \zref \right)
\end{align}
where $\Delta\chi = \chi(z)-\chi(\zref)$.
We assume $\xi$ varies much more quickly than $\pu$ or $\bu$.  Since $\xi$ is non-zero only over
a small redshift range $z\sim \zref$, we arrive at
\begin{align}
w(\theta, \zref) = \ & \pu(\zref) \bu(\theta, \zref) \bref(\theta, \zref) \ \times \nonumber \\ & \int dz\ \xi\left( \left[ \Delta \chi^2 + \chi(\zref)^2 \theta^2 \right]^{1/2} ; \zref \right) .
\end{align}
It is useful to rephrase the correlation function in terms of the physical transverse
separation $r=D_A\theta = (1+z)^{-1}\chi\theta$.
The integral of the correlation function is simply the projected correlation function
for matter,
\be
\wm(r, \zref) = \int dz\ \xi\left( \left[ (\chi(z)-\chi(\zref))^2 + (1+z)^2r^2 \right]^{1/2} ; \zref \right) .
\ee
Again, note we have opted to utilize {\it physical} units for $r$.
Plugging in we arrive at
\be
w(r, \zref) = \pu(\zref) \bu(r, \zref) \bref(r, \zref) \wm(r, \zref).
\ee

If we define the growth $G(r,\zref)$ relative to some arbitrary redshift $z_0$ via
\be
\wm(r,\zref) = G(r,\zref) \wm(r,z_0)
\ee
our final expression becomes
\be
w(r,\zref) = \pu(\zref) \bu(r, \zref) \bref(r, \zref) G(r,\zref) \wm(r,z_0).
\ee
Note $G$ is not necessarily the linear growth factor, and in fact, $G$ can depend on the length scale $r$.  We collect $\bu(r,\zref)$, $\bref(r,\zref)$, and $G(r,\zref)$ into a single source function $f(r,\zref)$ to write:
\be
w(r, \zref) = \pu(\zref) f(r,\zref) \wm(r)
\ee
where we have made the evaluation $z=z_0$ of $\wm$ implicit.  We will address how
we handle the function $f(r,\zref)$ momentarily.

We measure the angular correlation function by counting the number of pairs between our unknown and reference data $\Du\Dr$ separated over a range of scales $\Rmin$ to $\Rmax$ weighted by some function $W(r)$.  We take
$W(r)$ to be a power-law, $W(r)\propto r^\alpha$.
We discuss how we select the value $\alpha$ and the radial range $\Rmin$
to $\Rmax$ in Sections~\ref{sec:sdss} and \ref{sec:des} below.

Given the true angular correlation $w$, the number of unknown--reference
pairs is:
\be
\Du\Dr(z) = n_{\mathrm{u}} n_{\mathrm{r}} A_{\mathrm{survey}} A_{\mathrm{shell}} \int_{\Rmin}^{\Rmax} dr\ W(r) \left[1+w(r, z)\right] \ ,
\ee
where $n_{\mathrm{u}}$ and $n_{\mathrm{r}}$ are the number densities of the unknown and reference samples, $A_{\mathrm{survey}}$ is the area of the survey, and $A_{\mathrm{shell}}$ is the area of the shell over which the correlation function is computed.
Similar expressions hold for random point combinations $\Ru\Rr$, $\Du\Rr$, and $\Ru\Dr$,
only with $w=0$.  It is easy to check that the \citet{Landy:1993aa} estimator with
these pair counts results in an estimate of the weighted average correlation function
$\hat{w}$. 
Of course, one could compute the correlation function in narrow bins first,
and then average, but the two procedures are not equivalent, and we expect weighting
the pairs before averaging results in a more stable estimator since the averaging is
done before one takes the ratio of the data.  Our estimator for the weighted average correlation
function is
\begin{equation}
    \hat{w}(z) =
\frac{\DD(z) \NRR}{\RR(z) \NDD}
- \frac{\DRu(z) N_{\Ru}}{\RR(z) N_{\Du}}
- \frac{\DRr(z) N_{\Rr}}{\RR(z) N_{\Dr}}
+ 1
\label{eq:LS}
\end{equation}
Where $N_{\Du}$ is the number of unknown objects, and similar definitions hold for $N_{\Dr}, N_{\Ru}, N_{\Rr}$.

In practice, the \ngmix\ selection is not uniform in space,
and its spatial structure has not been characterized.  Consequently, a random catalog for the \ngmix\ galaxies does not exist.  When the unknown
sample does not have a well-characterized random catalog, we rely
instead on the estimator
\begin{equation}
\hat{w}_{\mathrm{noRu}}(z) = \frac{\DD(z) N_{\Rr}}{\DRu(z) N_{\Dr}} - 1 \ .
\label{eq:noRu}
\end{equation}

The expectation value of both estimators in the limit of infinitely large random catalogs is
\be
\avg{\hat{w}(z)} = \pu(z) \frac{\int_{\Rmin}^{\Rmax} dr\ W(r) \ f(r,z) \wm(r)}{\int_{\Rmin}^{\Rmax} dr\ W(r)}
\label{eq:hatw}
\ee
In practice, the finite number of random points must introduce second order corrections to
our estimator.  However, such corrections do not appear to have any significant effect on our
results given the large number of randoms we use ($R/D\gtrsim 100$).

We define the function $f(z)$ via
\be
f(z) = \frac{ \int_{\Rmin}^{\Rmax} dr\ W(r) \ f(r,z) \wm(r) }{ \int_{\Rmin}^{\Rmax} dr\ W(r) 
}
\ee
With this definition, the expectation value for $\hat w$ becomes
\be 
\avg{\hat{w}(z)} = \pu(z) f(z)
\ee 
where $f(z)$ is an unknown function that characterizes the (possibly non-linear) growth
in the correlation function, and/or possibly evolving non-linearities
in the correlation function.  We adopt a simple power-law parameterization for $f(z)$,
\be
f(z) = f_0 (1+z)^{\gamma}.
\label{eq:biasevolution}
\ee

With this parameterization, our final expression for the expectation value of the cross-correlation is
\be 
\avg{\hat{w}(z)} = \pu(z) f_0 (1+z)^\gamma
\ee 

This expression has two parameters: $\gamma$, which characterizes the redshift
evolution in the correlation function (including possible non-linear effects, and $f_0$,
which characterizes primarily nuisance deviations from unity normalization over the range of redshifts sampled by the reference sample.

Given a model for the redshift distribution $\pu(z)$, which may itself
depend on unknown parameters, we can recover redshift distribution 
through the usual $\chi^2$ statistic, 
\begin{equation}
    \label{eq:bias_noshift}
    \chi^{2} = \Delta w \hat{\Sigma}^{-1} \Delta w \ ,
\end{equation}
where $\hat{\Sigma}$ is our estimated covariance in the observed cross-correlation $\hat w$, and where
\be 
\Delta w = \hat{w} - f_0 (1+z)^{\gamma} \phi(z) \ .
\label{eq:model_noshift}
\ee
In practice we find that parameterizing $f_0$ as $e^k$ and fitting instead for $k$ improves performance.

We calculate the pairs using the code
\texttt{TreeCorr},\footnote{\url{https://github.com/rmjarvis/TreeCorr}} a
\texttt{C} and \texttt{Python} package for efficiently computing 2-point and 3-point correlation
functions \citep{Jarvis:2004ab}.
  We estimate
the covariance matrix of our estimation of $\phi_{\rm{u}}$ with 100 jackknife regions
on the survey footprint.

\section{Testing the Cross-Correlation Method with the Sloan Digital Sky Survey}
\label{sec:sdss}

The simplest way to test whether we can use \redmapper\ clusters to measure redshift 
distributions through \cczs\ is to
look at the performance of \cczs\ on spectroscopic galaxies.
We approach this problem in two steps:
first, we verify the validity of \cczs\ by determining the redshift
distribution of a spectroscopic sub-sample of galaxies by cross-correlating
it against an independent spectroscopic reference sample.
Then we examine the recovery of the redshift
distribution of \sdss\ galaxies using \redmapper\ clusters as the
reference sample.

\begin{figure}
\begin{center}
\includegraphics[width=0.5 \textwidth]{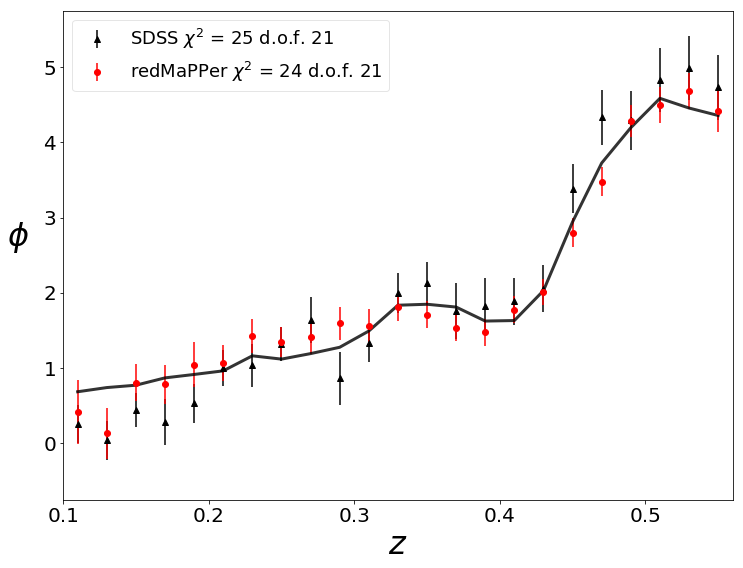}
\end{center}
\caption{A comparison of the \ccz\ recovery of the \sdss\ redshift distribution with their spectroscopic redshifts.
The solid line is the actual distribution of galaxies from
    spectroscopic data.  The black points show the recovered
    \sdss\ redshift distribution using \ccz\ with a spectroscopic
    sample of \sdss\ galaxies as a reference sample.  
    The red points show the corresponding redshift distribution
    when using \redmapper\ galaxy clusters as the reference sample. 
    Both sets of points account for the best-fit redshift evolution
    in the clustering bias.}
\label{fig:dndz_sdss}
\end{figure}

We combine the CMASS and LOWZ samples and randomly split the galaxies into a reference
sample containing 80\% of the galaxies and an unknown sample containing the
remaining 20\%. The random selection ensures that the redshift distributions of the samples are identical.
The reference sample is divided according to spectroscopic
redshift into bins of width $\delta z = 0.02$.
Pair counts of the sample and
randoms are counted as described above in Section~\ref{sec:methods}. The
pair counts are integrated into a single scalar from 100 kpc to 10 Mpc and
weighted by $g(R)=R^{-1}$. In order to facilitate
later comparison with \redmapper\ clusters, which have a more limited redshift
range than the spectroscopic sample, we normalize the distribution such that its integral from $z=0.1$ to $0.56$ is one. The highest redshift is set by the highest redshift for which
we have enough \redmapper\ clusters to obtain reasonable cross-correlation measurements.
Covariances are estimated from
jackknife samples over the survey footprint.
We find that ignoring the bias evolution (setting $\gamma = 0$) produces a poor fit between the clustering and spectroscopic measurements ($\chi^2/\mathrm{dof} = 58 / 23$). We fit the bias evolution model described by Equation~\eqref{eq:model_noshift} and find that this brings our results into agreement ($\chi^2/\mathrm{dof} = 25 / 21$), with $\gamma = -1.7 \pm 0.5$.
The results can be seen in
the black triangles of Figure~\ref{fig:dndz_sdss}.
These results validate our cross-correlation technique.

Next we repeat the same exercise, only now the reference objects are the \redmapper\ clusters, and the unknown objects are the entire
\sdss\ CMASS and LOWZ samples. We emphasize that our goal here is to test whether we can use \redmapper\ clusters to measure redshift distributions through the \ccz, even if our signal includes scales well within the radius of a \redmapper\ cluster.
We again fit the same model to the clustering signal, and find that the \ccz\ method with \redmapper\ clusters as a reference sample
successfully recovers the correct redshift distribution of the unknown sample within noise
($\chi^2/\mathrm{dof} = 24 / 21$), with $\gamma = -0.2 \pm 0.4$.
The
results are shown as red circles in Figure~\ref{fig:dndz_sdss}.
These results validate the idea that replacing spectroscopic reference samples with high-quality photometric reference samples is a viable 
approach to calibrating photometric redshift distributions via
cross-correlations.

\begin{figure*}
\begin{center}
	\includegraphics[width=\textwidth]{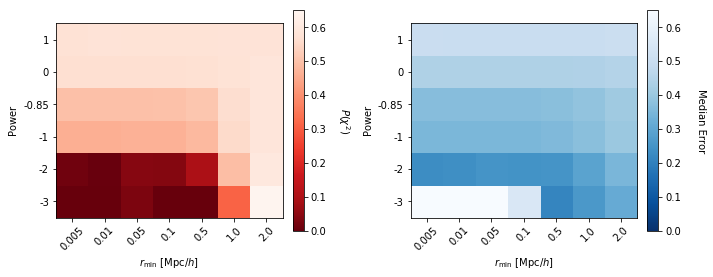}
\end{center}
\caption{Effect of varying inner integration range $r_{\mathrm{min}}$ and
weight power $\alpha$ on fit $\chi^{2}$ (Left) and median error (Right) for the
recovery of the \sdss\ redshift distribution using \sdss\ galaxies. Setting the
weight power too high will result in larger errors by emphasizing the weaker signal at larger scales, while setting the
power too low will bias the measurement as extra weight is given to the non-linear
regime. The power $\alpha=-0.85$ is included as a row because it is the slope of a powerlaw
fit to the cross-correlations.}
\label{fig:params_sdss}
\end{figure*}
\begin{figure*}
\begin{center}
     \includegraphics[width=\textwidth]{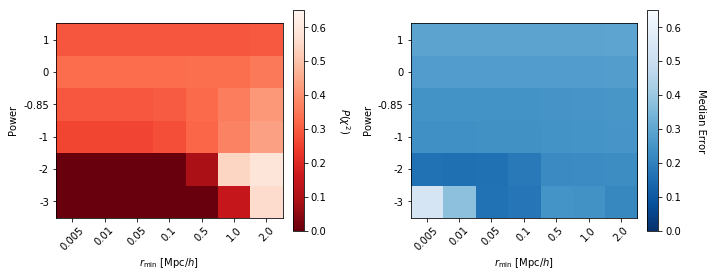}
\end{center}
\caption{Effect of varying inner integration range $r_{\mathrm{min}}$ and
    weight power $\alpha$ on fit $\chi^{2}$ (Left) and median error (Right) for
    the recovery of the \sdss\ redshift distribution using \redmapper\ galaxy
    clusters. As with Figure~\ref{fig:params_sdss}, the particular choice of
    hyperparameters can swing the recovered distribution's fit and error
    properties significantly.}
\label{fig:params_redmapper}
\end{figure*}

\Cczs\ convert angular clustering signal into estimates of the redshift distribution, but the error in clustering estimates varies with angular scale due to sample variance, shot noise, and survey systematics. Consequently, we test how our choice of weighting function affects the
recovered redshift distribution, and how varying the radial range, particularly
the inner radius, impacts our results. We consider simple power-law weighting
functions $g(r)=r^{\alpha}$. We expect \textit{a priori} that
the optimal power will be close to the approximate power law exponent of
the correlation function itself, $\alpha \sim -0.85$.
Given this expectation, we expect the primary sensitivity to the radial range
$r\in[\Rmin,\Rmax]$ will be through the scale $\Rmin$, which is why we focus on
$\Rmin$ in this study.

We find that there is a trade-off between bias and variance:
varying the power of the weight function and the inner radius of the angular
integration can significantly decrease the estimated covariance at the cost of
noticeable biasing.
Figures~\ref{fig:params_sdss}-\ref{fig:params_redmapper} summarize
this trade-off. Each Figure has two colored grids showing the effect of
varying the power of the weight function and the inner angular integration
range on the recovered redshift distribution. In the left panel, we take
the overall best $\chi^2$ fit and plot the probability $P_{k}(\chi^2)$
that $\chi^2$ be larger than the observed value given the degrees of freedom
in the analysis.  When this probability is small, our recovered redshift distribution
is not an acceptable fit to the data.
In the right panel, we plot the median error of the recovered redshift distribution as estimated from jackknife samples, $\textrm{Median} \left( \sigma_{\phi} \right)$.
For both the \sdss\ spectroscopic galaxy reference sample and the \redmapper\ reference
sample we find that $\Rmin=0.1\ \Mpc$ and $\alpha=-1$ provide a nearly optimal tradeoff
between accuracy and precision. However, we note that the specific response to variations in these parameters may differ in other samples.

\section{Application to 
the Dark Energy Survey: Combining \PHOTOZS\ with
Cross-Correlation Methods}
\label{sec:des}

It is vital for \des\ to accurately constrain the redshift distributions of
objects that are in tomographic bins for measurements of cosmic shear or baryon acoustic
oscillations.
Traditionally, one places galaxies into tomographic bins based on
their photometrically-estimated probability distribution $P(z)$,
and then one ``stacks'' these $P(z)$'s to estimate the redshift
distribution of sources in a bin.
The \ccz\ method as implemented here does not provide insight
into which objects should go into each tomographic bin, but it
does constrain the redshift distribution
of the objects in each bin.  Consequently, it can provide a critical
systematics cross-check for
\photoz\ methods. This is especially desirable because the inputs and
systematics that affect the
\ccz\ method are largely independent of those affecting
\photoz\ methods.
In particular, while spectroscopic redshift incompleteness is the primary difficulty
affecting \photoz\ algorithms, this systematic is completely irrelevant for \ccz.

In this Section we implement a \ccz\ measurement of the redshift distribution of
\ngmix\ galaxies using \redmapper\ clusters as the reference sample.
The galaxies are placed into three tomographic bins with
edges at $0.3, 0.55, 0.83, 1.3$ according to
the mean redshift of the \texttt{SkyNet} \photoz\ code. While the
\ngmix\ galaxies span a wide range of redshifts, we must limit our analyses to
$0.2 < z < 0.84$, where we have \redmapper\ clusters with reliable redshifts. Because of this limited redshift range, we do not perform the \ccz\ measurement on the third tomographic bin.

We repeat the same sort of \ccz\ analysis as in Section~\ref{sec:sdss}, where
we use \redmapper\ clusters to measure the redshift distribution.
We set the power of the weighting term, $\alpha$ to -1 and integrate from 100 kpc to 10 Mpc.
As with \sdss, we begin with the simple model in which $\gamma=0$, and the
cross-correlation function is directly proportional to (and therefore an estimator
of) $\phi(z)$.

\begin{figure*}
    \begin{center}
         \includegraphics[width=\textwidth]{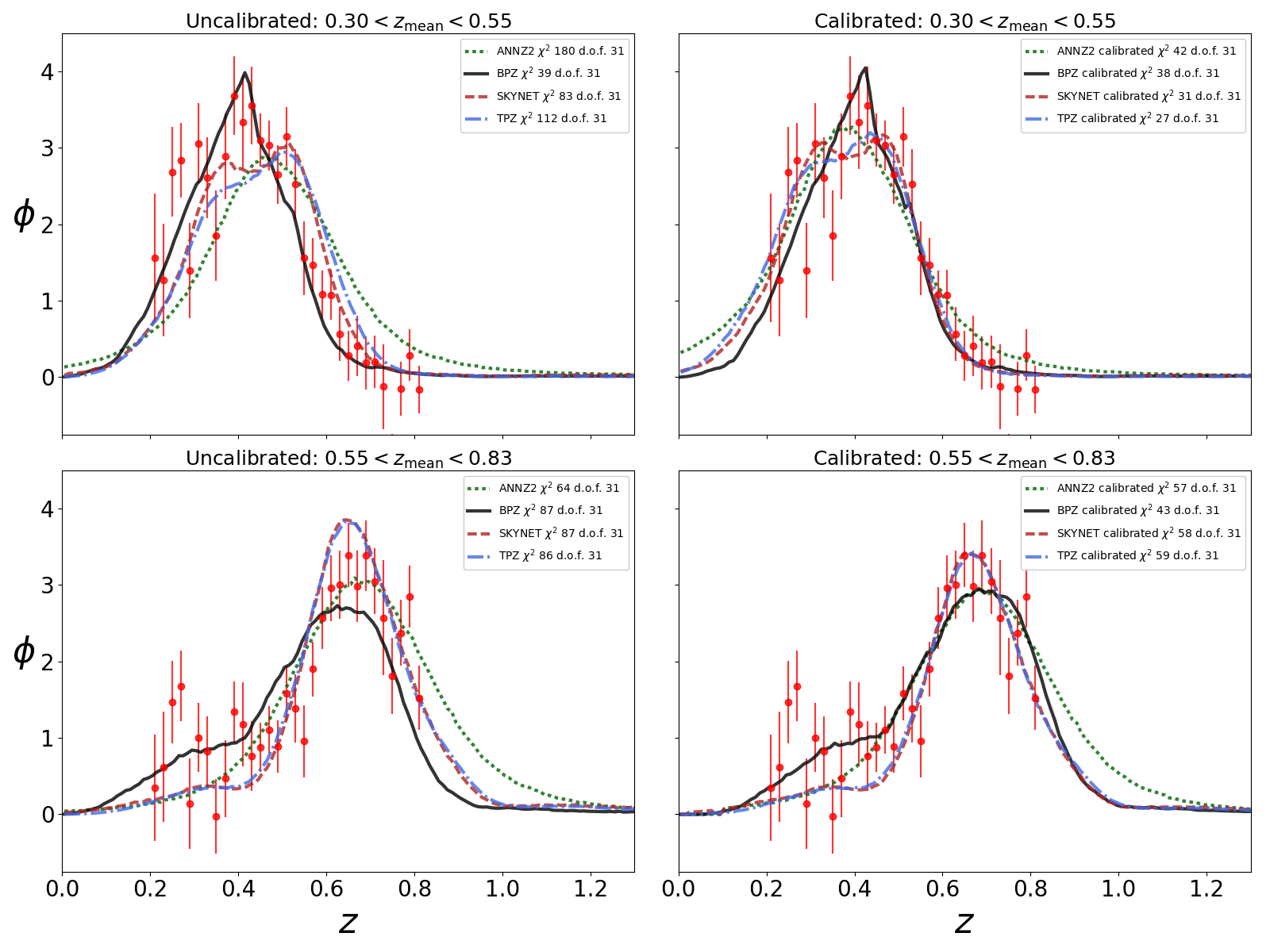}
    \end{center}
    \caption{Redshift distribution of \ngmix\ galaxies using \redmapper\
        galaxy clusters before (left) and after (right) the calibration of 	Equations~\eqref{eq:bias} and~\eqref{eq:model}.
        Lines represent different \photoz\ estimation codes,
        while the points represent the \ccz\ method. Each row of panels
        represent a different tomographic bin, with galaxies selected
        according to the \texttt{SkyNet} \photoz\ method.
        Redshift distributions are normalized to one over the range of the \ccz\ sample $0.2 < z < 0.84$.
        The correction values are shown in
    Table~\ref{tab:des_sv_bias_noru}.
    Note that in practice each \photoz\ code modifies the predicted redshift distribution
    from the cross-correlation analysis through the bias evolution term.
    In order to make it easier to visually compare the different \photoz\ codes, we apply the corrections associated with \cczs\ to the \photoz\ redshift distributions, which allows
    a direct comparison with the cross-correlation data for all \photoz\ codes.}
    \label{fig:dndz_sv_noru}
\end{figure*}

We present the results of these
analyses in the left column of Figure~\ref{fig:dndz_sv_noru}.
Among the four codes we consider, only one has an acceptable $\chi^2$, namely \texttt{BPZ}.
For the second tomographic bin, none of the \photoz\ codes have an acceptable $\chi^2$. The $\chi^2$ of the predicted \photoz\ distributions relative to
the recovered redshift distribution are summarized in Table~\ref{tab:des_sv_bias_noru}.

We seek to reconcile \photoz\ and \ccz\ data where possible. 
As discussed in \citet{Bonnett:2015aa}, the primary source
of systematic uncertainty affecting the \photoz\ distributions is an overall photometric
redshift bias.  We utilize our \ccz\ data to calibrate this photometric redshift bias.  In practice, this means replacing the
photometric redshift distribution $\pu(z)$ by
a the function $\pu(z-\Delta z)$ where $\Delta z$ is the
photometric redshift bias of the algorithm in question. 
We write
\begin{equation}
    \label{eq:bias}
    \chi^{2} = \Delta w \hat{\Sigma}^{-1} \Delta w
\end{equation}
where
\be 
\Delta w = \hat{w} - e^{k} \left(\frac{1 + z}{1 + z_0} \right)^{\gamma} \phi(z - \Delta z)
\label{eq:model}
\ee
The quantity $k$ accounts for an overall relative normalization of the distributions,
since we normalize $\hat{w}(z)$ to unity within the redshift range
$[0.2,0.84]$, whereas $\phi(z)$ is normalized to unity within $[0,\infty]$.
We also account for clustering bias evolution from Equation~\eqref{eq:biasevolution} via our parameterization of $(1+z)^{\gamma}$, which is normalized by $1 + z_0$, where $z_0$ is chosen to be 0.52, the center of the cross-correlation redshift range. 
Because redshift biases
and normalizations may change across tomographic bins, we define separate variables
$k_i, \Delta z_{i}$ for each tomographic bin. 
In contrast, we force each tomographic bin to simultaneously fit the same $\gamma$.  This is consistent with our choice to encode all redshift evolution
information into $\gamma$, although we note that clustering bias evolution may be induced by the \photoz\ selection of the \ngmix\ galaxies into different tomographic bins.  Not surprisingly, allowing for independent $\gamma$ in each of the tomographic
bins can introduce large uncertainties in the recovered distributions since any individual tomographic
bin is too narrow to properly constrain 
the redshift-dependent evolution of the clustering bias; one really needs the full range of redshifts probed by the data.

We also utilize our prior knowledge of the redshift distributions from the
\photoz\ codes.
We model our prior on the redshift bias $\Delta z_{i}$ for each tomographic
bin as a Gaussian centered on 0 with width $\sigma = 0.05$ from prior analyses of \photoz\ uncertainties from \citet{Bonnett:2015aa}. Thus, the expression
we maximize is:
\begin{equation}
    \label{eq:prior}
    \ln \mathcal{L} = -\frac{1}{2} \chi^{2} \left(k_i, \Delta z_i, \gamma \right) - \sum_{j} \frac{1}{2} \frac{\Delta z_j^2}{2\times 0.05^2}
\end{equation}
When \cczs\ are able to strongly constrain the redshift bias
(as is the case here),
the particular choice of prior does not matter.
In cases when \cczs\ are unable to constrain the redshift bias
we simply recover our prior. By using priors informed by \photoz\ methods we open an avenue through which we may combine \photoz\ and \ccz\ measurements.

We sample Equation~\eqref{eq:prior}
using \texttt{emcee},\footnote{\url{http://dan.iel.fm/emcee}} an affine
invariant Markov Chain Monte Carlo ensemble sampler
\citep{Foreman-Mackey:2013aa}. Results of these fits are presented in
the right column of Figure~\ref{fig:dndz_sv_noru} and in Table~\ref{tab:des_sv_bias_noru}. 
Allowing for an overall redshift bias, all four \photoz\ codes have an acceptable
$\chi^2$, with \texttt{TPZ} resulting in the best fit.  The good agreement between
the codes is also apparent from Figure~\ref{fig:dndz_sv_noru}.
By contrast, only one of the codes, \texttt{BPZ}, has an acceptable $\chi^2$
for the second tomographic bin.  Figure~\ref{fig:dndz_sv_noru} makes it apparent
that most codes agree quite well on the shape of the \photoz\ distribution
in the vicinity of the average redshift of the galaxies in the bin, but differ
significantly on the amplitude of a low-redshift tail in the distribution.  
The relatively small area of the \sv\ region makes the correlation function
measurements themselves very noisy in this regime, though the data does seem
to indicate a preference for a larger tail, in agreement with the \texttt{BPZ}
prediction.  Future analyses from the full \des\ footprint should be able to
clearly resolve this feature, and establish whether the ``blips'' seen
in the lower right-hand panel of Figure~\ref{fig:dndz_sv_noru} are real 
or just random fluctuations.

It is important to note that we also find discrepancies in our constraints on $\gamma\ $ from the different \photoz\ codes.  This difference is easy to explain: the shape $\phi(z)$ of the distribution is clearly degenerate with $\gamma$.  Since the different \photoz\ codes
have different shapes, $\gamma$ takes on a different value for each \photoz\ code. 
This demonstrates that
the shape of $\phi(z)$ estimated from the traditional \photoz\ method
constitutes an important systematic for our proposed method.  Future work that seeks
to implement the proposed method in cosmological analyses must properly quantify the
associated systematic error, for instance through the use of simulated data.

\begin{table*}
\begin{center}
\begin{tabular}{| l | l || r | r | r | r | r |}
  \hline
  \rule{0pt}{4ex}    
   Photo-$z$ & Bin & $\chi^{2}_{\mathrm{raw}, \mathrm{noRu}}$ & $\chi^{2}_{\mathrm{cal}, \mathrm{noRu}}$ & $\mathrm{dof}$ & $\Delta z_{\mathrm{noRu}}$ & $\gamma_{\mathrm{noRu}}$  \\
  \hline
  & &  \\
ANNZ2 	& 1	 	& $180$ & $42$ 	& $31$ 	& $0.082 \pm 0.011$ 	& $0.6 \pm 0.5$  \\ 
ANNZ2 	& 2	 	& $64$ 	& $57$	& $31$ 	& $-0.020 \pm 0.014$ 	& $0.6 \pm 0.5$  \\ 
BPZ 	& 1	 	& $39$ 	& $38$	& $31$ 	& $-0.005 \pm 0.006$ 	& $0.1 \pm 0.4$  \\ 
BPZ 	& 2	 	& $87$ 	& $43$	& $31$ 	& $-0.058 \pm 0.011$ 	& $0.1 \pm 0.4$ \\ 
SKYNET 	& 1	 	& $83$ 	& $31$	& $31$ 	& $0.045 \pm 0.008$ 	& $0.5 \pm 0.5$  \\ 
SKYNET 	& 2	 	& $87$ 	& $58$	& $31$ 	& $-0.018 \pm 0.007$ 	& $0.5 \pm 0.5$  \\ 
TPZ 	& 1	 	& $112$ & $27$ 	& $31$ 	& $0.061 \pm 0.009$ 	& $0.5 \pm 0.5$  \\ 
TPZ 	& 2	 	& $86$ 	& $59$ 	& $31$ 	& $-0.019 \pm 0.008$ 	& $0.5 \pm 0.5$  \\ 
                 & &  \\
  \hline
\end{tabular}
\caption{
Table of results from fitting \photoz\ codes to \ccz\
analyses from \redmapper\ clusters with \sv\ \ngmix\ galaxies by maximizing
Equation~\eqref{eq:prior}. The `raw' $\chi^2$ is the goodness of fit before allowing
for an overall photometric bias of the \photoz\ codes.  The `cal' $\chi^2$
value is that obtained after calibrating this bias using cross-correlations. 
}
\label{tab:des_sv_bias_noru}
\end{center}
\end{table*}

Keeping in mind the above important caveat, it is reassuring to see 
that for the majority of the \photoz\ codes the posterior on the redshift bias 
is less than 0.05 in magnitude, consistent with the expectations 
of \citet{Bonnett:2015aa}.  In this context, it is also worth noting
that our posteriors on $\Delta z$ measure 
the \textit{relative} redshift bias between our \textit{unknown} and \textit{reference} samples. \redmapper\ itself has \photoz\ redshift biases that are unconstrained at the level of $\sigma_{\Delta z} \sim 0.003$. These are currently significantly smaller than the uncertainties from \cczs , but may need to be accounted for as the statistics of \cczs\ improve to \yearfive\ levels.

Nevertheless, it is encouraging to see that the 
statistical precision from this analysis in the \sv\ region ($\sim 139 \ \mathrm{deg}^2$) is 
not far from what would be required for a \yearfive\ ($\sim 5000 \ \mathrm{deg}^2$)
cosmology analysis.  
Of course, in practice, detailed simulation work will be required to properly 
characterize the systematic uncertainties associated with this type of analysis,
and possibly motivate alterations to the simple algorithm proposed here. 
Motivated by these results, we have launched such a study in preparation
for future \des\ cosmological analyses.

As a final test of our proposed analysis,
we revisit the \sdss\ data set, and test our full algorithm
there by calibrating the relative redshift bias between \sdss\ spectroscopic
galaxies and \redmapper\ clusters. 
As expected, we find the recovered redshift bias is consistent with zero
$\Delta z = (-5.1 \pm 4.6) \times 10^{-3}$, while the clustering bias evolution parameter $\gamma = -0.3 \pm 0.5$. This level of uncertainty is consistent with \it a priori \rm 
expectations of the \redmapper\ redshift bias.

\section{Cosmological Forecasts}
\label{sec:cosmo}

\begin{figure}
\begin{center}
     \includegraphics[width=0.50 \textwidth]{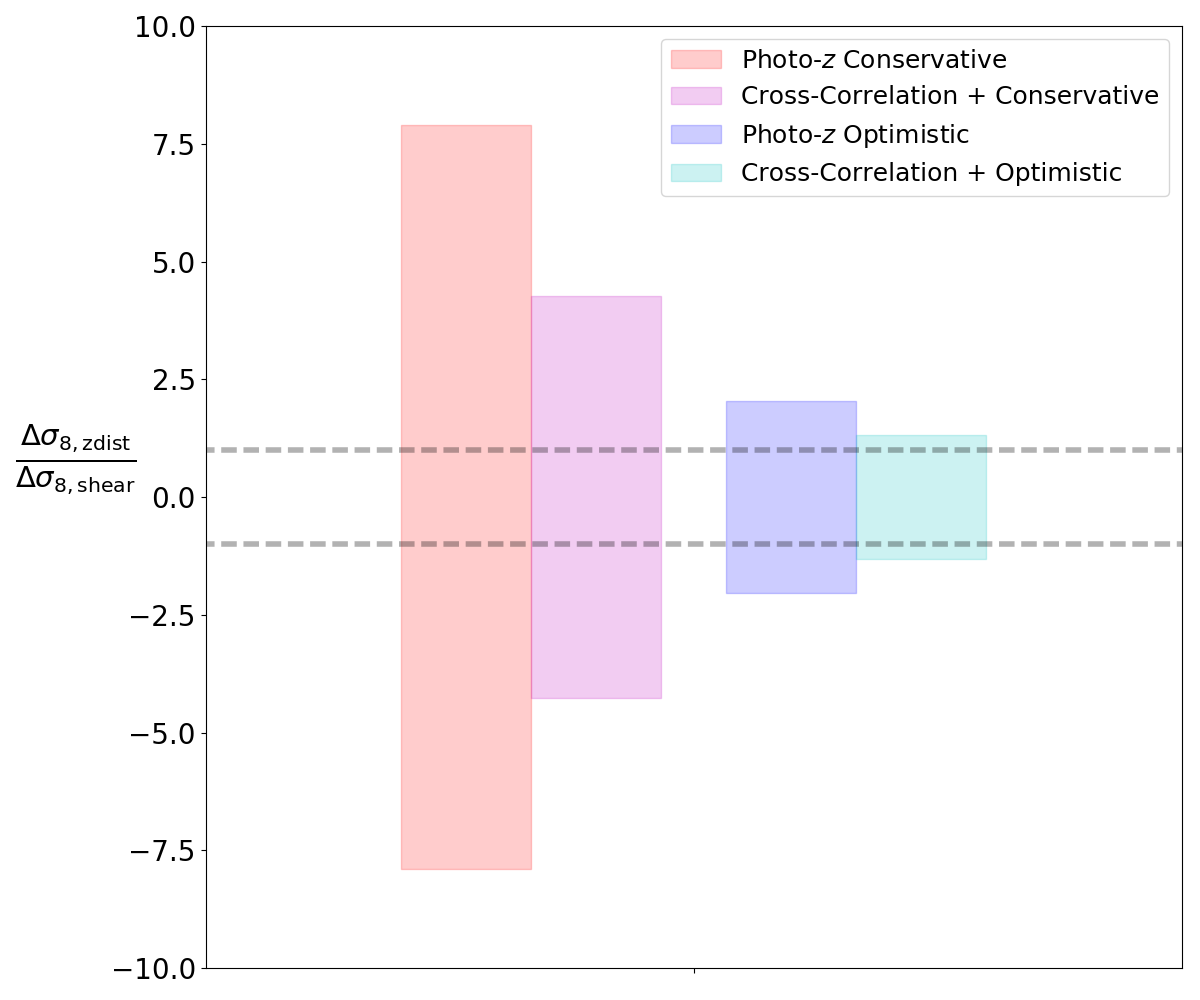}
\end{center}
\caption{Ratio of the variation in $\sigma_8$ due to forecasted \yearfive\ uncertainties from redshift distributions to shear statistical uncertainties. When this ratio is less than one, the uncertainty on $\sigma_8$ is dominated by errors in shape measurements; when the ratio is greater than one, errors from the redshift distribution dominate.
The red (blue) band is the uncertainty due to the \texttt{SkyNet} \photoz\ code with the conservative (optimistic) uncertainty $\sigma_{\Delta z} = 0.05$ ($\sigma_{\Delta z} = 0.02$) in the bias of the redshift distribution.
The magenta (cyan) band is the residual uncertainty from using \cczs\ to constrain the redshift bias $\Delta z$. The grey dashed lines are guidelines for $\pm 1$.
Tomographic constraints are limited by the lack of overlap with \redmapper\ clusters in the third (highest redshift) tomographic bin,
which prevents any \ccz\ calibration in that bin.
}
\label{fig:dndz_cosmo}
\end{figure}

We now estimate the efficacy of \ccz\ redshift calibration for cosmological analyses of
\yearfive\ cosmic shear data. We start with the \sv\ \ngmix\ cosmic shear two-point correlations.
In order to perform our forecast, we must also specify a fiducial
redshift distribution of sources. 
In keeping with the cosmic shear analysis of \citet{Becker:2015aa}, we take \texttt{SkyNet} as our fiducial redshift distribution.
Here, we focus on the recovery of $\sigma_8$ (holding all other cosmological
parameters fixed to their fiducial value) through
the cosmology analysis pipeline used in
\citet{Bonnett:2015aa}.\footnote{\url{https://github.com/matroxel/destest}}

We focus on two key sources of uncertainty in cosmic shear measurements:
errors in the
measurement of the shear two point correlation function, and errors in our
characterization of the redshift distribution from either \photoz\ or \ccz\
systematics.
We compare the posteriors in $\sigma_8$ for three different fiducial analyses.  First, an
analysis that includes only shape noise, and the source redshifts are assumed to be perfectly known.
Second, an analysis that assumes no shape noise, but considers only the error in $\sigma_8$ introduced
by an overall unknown \photoz\ redshift bias with current \photoz\ priors.
Third, an analysis that assumes no shape noise, but considers
the error in $\sigma_8$ introduced by unknown \photoz\ redshift bias that is calibrated using \ccz.

To model our \yearfive\ shear covariances, we scale the \sv\ cosmic shear
covariances by the area of the final survey, so that we divide the \sv\ cosmic
shear covariance matrix by $5000 / 139$, the ratio of \yearfive\ to \sv\ SPTE survey areas. Under the assumption that the galaxy density measured in \textit{SV} will be equal to what we will have in Y5, this scaling is the improvement in statistical power from repeating our \textit{SV} analysis over the full Y5 footprint.

To estimate the uncertainty in the recovered $\sigma_8$ due to an unknown
photometric redshift bias with some prior and no cross-correlation data,
we sample the redshift
distributions $P(z - \Delta z)$ for the source galaxies, where $\Delta z$ 
is the redshift bias.  We find the best fit $\sigma_8$ given the stacked $dN/dz$, and repeat
our measurement 250 times, randomly drawing the redshift bias from the prior $\Delta z = 0.00\pm 0.05$. This is the level of constraint if the same \photoz\ algorithms from \sv\ were used in \yearfive, and is hence our `conservative' case.
The standard deviation in the recovered $\sigma_8$ values is the uncertainty in $\sigma_8$ due
to photometric redshift uncertainties.
We also consider the case where
\photoz\ algorithms improve in the measurement of systematic redshift bias to
$\sigma_{\Delta z} = 0.02$.  We refer this as our `optimistic' scenario.
Note `optimistic' should not be confused with `unrealistic'; we expect future
\photoz\ codes to be able to reach this level of accuracy and precision. 

To measure the effect of \ccz\ on the recovery of $\sigma_8$, we begin by
estimating the uncertainty in the photometric redshift bias $\sigma_{\Delta z}$
as estimated from the \ccz\ method, as per our discussion in
Section~\ref{sec:des}.  This constraint depends on our fit for
Equation~\eqref{eq:bias}. We scale our estimated errors $\hat{\Sigma}$ by the
ratio of the survey areas between the \yearfive\ and \sv\ surveys (i.e.
$5000/139$) to simulate the increase in signal. By then setting $\hat{\phi} =
\phi$ and fitting as described in Section~\ref{sec:des}, we sample the space of
solutions that are consistent with the \ccz\ method. These samples provide the
redshift bias $\Delta z$ that goes into our cosmology analysis. This is roughly
equivalent to the simpler method of scaling the errors in the photometric
redshift biases in Table~\ref{tab:des_sv_bias_noru} by the square root of the ratio
of the survey areas between the \yearfive\ and \sv\ surveys, but has the
capability for incorporating more complex modifications of the redshift
distribution in the future.  When comparing the optimistic and conservative cases, we use
the fact that we know from the \photoz\ algorithms that $\sigma_{\Delta z} = (0.02, 0.05)$ respectively by
putting that information into our prior on the redshift bias via
Equation~\eqref{eq:prior}. 
This makes a significant difference because the
\ccz\ cannot calibrate the highest tomographic bin due to a lack of overlap with the \redmapper\ clusters, and so we revert back to the prior there.
In addition, using the sampling method used to characterize the cosmological uncertainties
from \photoz, we also add a systematic redshift error  $\sigma_{\Delta z} =
0.003$ to our reference sample to account for the 
photometric redshift bias expected for \redmapper\ clusters. 

Figure~\ref{fig:dndz_cosmo} compares the errors in $\sigma_8$ from each of the three
analyses considered above. We plot the ratio of the uncertainty on $\sigma_8$ arising from uncertainty in the mean of the redshift distribution (either from \photoz\ or \ccz\ calibration) to the uncertainty on $\sigma_8$ arising from shape noise in the shear two point correlation function. When this ratio is less than one, the uncertainty on $\sigma_8$ is dominated by errors in shape measurements; when the ratio is greater than one, errors from the redshift distribution dominate.
There is a noticeable improvement in the recovered cosmological parameters in our conservative analysis, but 
only a modest improvement in our optimistic scenario.  This largely reflects the fact that our analyses 
does not constrain the highest tomographic redshift bin, which contains a large fraction of the cosmological
signal. Nevertheless, it is very encouraging that even with this simple approach, and without incorporating
the gains from high redshift cross-correlation analyses that may be enabled in the future, a combined
\photoz\ and cross-correlation analysis appears to have the potential to control photometric redshift
systematics.  
Consequently, our results provide strong motivation for further investigations
into the possible use
of photometric cross-correlations for \photoz\ calibration.

\section{Summary and Conclusions}
\label{sec:conclusions}

We examined the performance of \ccz\ methods when photometric redshifts are
used instead of spectroscopic galaxies as a reference sample.
We first verified that \ccz\ methods work with spectroscopic reference samples
by looking at the \ccz\ of \sdss\ galaxies. We then found that \redmapper\
galaxy clusters, despite having photometric redshifts, are also viable
reference samples for measuring accurate redshift distributions.

Having validated our methodology on \sdss\ spectroscopic galaxies, we turned toward
the \des\ \ngmix\ sample in the \sv\ data set. There, we applied the \ccz\ method and compared
our measured redshift distributions to those measured by a variety of \photoz\
codes. We developed a formulation for calibrating systematic redshift biases
in the \photoz\ codes, using information from both the \photoz\ and \ccz\ methods to improve the characterization of redshift distributions. We found that the recovered redshift bias relative to the \ccz\ measurement is typically less than
$\sim 0.05$, with the posterior of the redshift bias being uncertain at the
$\pm 0.01$ level. This level of uncertainty is comparable to the
required \yearfive\ uncertainty and should only improve with five more years of
data. However, systematic errors, particularly from the evolution of clustering bias in both the source and reference samples, as well as from discrepancies between the shapes of the \photoz\ and \ccz\ redshift distributions, need to be fully characterized for this method to be 
integrated into future cosmological analyses.  

Finally, we extrapolated the recovered statistical uncertainties to \yearfive\
data, and compared the statistical uncertainty in $\sigma_8$ from shape noise,
the photometric uncertainties from a fiducial \photoz\ algorithm with an
unknown redshift bias at the $2\%$ level, and a fiducial analysis that includes
calibrations from \cczs. The impact of the \ccz\ analysis depends on the
uncertainty that traditional \photoz\ algorithm can reach in the near future.
Even in our optimistic scenario, however, \cczs\ provide a non-negligible
improvement in the recovered cosmological constraints, while simultaneously
providing a critical consistency test of the \photoz\ calibration. 

These results firmly establish the feasibility of using photometric samples
as reference samples for calibrating photometric redshift distributions via \ccz,
and provides strong motivation for more detailed simulation-based studies
aimed at fully characterizing the systematic uncertainties in these methods.
In particular, we have already seen hints that systematic uncertainties in the
shape of the $\phi(z)$ estimated from traditional \photoz\ methods 
can introduce systematics in the recovered photometric redshift bias.
We are currently characterizing these systematics via numerical simulations for the \des\ Year 1 cosmology analysis (Gatti et al, in prep.).
Similarly, it is not difficult to imagine modifications of the methodology
adopted here that is more ideally suited to cosmological analyses. 
For example, cosmic shear analyses are primarily sensitive to \photoz\ uncertainties
through the photometric redshift bias and not the shape of the redshift distribution.  The methods presented here could be tailored towards these goals by matching the means of the distributions while down-weighting the tails of the
redshift distributions, thereby minimizing systematics associated with the detailed shape
of $\phi(z)$. We are also exploring alternative high fidelity photometric reference samples such as the luminous red galaxy \redmagic\ sample described in \citet{Rozo:2015aa}, which has considerably higher number density than \redmapper\ clusters (Cawthon et al, in prep.).
The fact that the \ccz\ provide a calibration tool that is demonstrably capable
of reaching the necessary statistical precision for Stage III dark energy experiments
is an important step forward in realizing the promise of the \des, and provides
strong motivation for further studies of this method.


\section*{Acknowledgements}
\label{sec:acknowledgements}
CPD would like to acknowledge Enrique Gaztanaga, Marco Gatti, Pauline Vielzeuf, Ross Cawthon, \`{A}lex Alarc\'{o}n for many fruitful conversations. He would also like to thank Ben Hoyle, Huan Lin, Ramon Miquel, and Michael Troxel for their many suggestions, which tremendously improved this paper. This work is partially supported by the Northern California Chapter of the ARCS Foundation, as well as by the U.S. Department of Energy under contract number DE-AC02-76-SF00515.
ER is supported by DOE grant DE-SC0015975 and by the Sloan Foundation, grant FG-2016-6443.

Funding for the DES Projects has been provided by the U.S. Department of Energy, the U.S. National Science Foundation, the Ministry of Science and Education of Spain, 
the Science and Technology Facilities Council of the United Kingdom, the Higher Education Funding Council for England, the National Center for Supercomputing 
Applications at the University of Illinois at Urbana-Champaign, the Kavli Institute of Cosmological Physics at the University of Chicago, 
the Center for Cosmology and Astro-Particle Physics at the Ohio State University,
the Mitchell Institute for Fundamental Physics and Astronomy at Texas A\&M University, Financiadora de Estudos e Projetos, 
Funda{\c c}{\~a}o Carlos Chagas Filho de Amparo {\`a} Pesquisa do Estado do Rio de Janeiro, Conselho Nacional de Desenvolvimento Cient{\'i}fico e Tecnol{\'o}gico and 
the Minist{\'e}rio da Ci{\^e}ncia, Tecnologia e Inova{\c c}{\~a}o, the Deutsche Forschungsgemeinschaft and the Collaborating Institutions in the Dark Energy Survey. 

The Collaborating Institutions are Argonne National Laboratory, the University of California at Santa Cruz, the University of Cambridge, Centro de Investigaciones Energ{\'e}ticas, 
Medioambientales y Tecnol{\'o}gicas-Madrid, the University of Chicago, University College London, the DES-Brazil Consortium, the University of Edinburgh, 
the Eidgen{\"o}ssische Technische Hochschule (ETH) Z{\"u}rich, 
Fermi National Accelerator Laboratory, the University of Illinois at Urbana-Champaign, the Institut de Ci{\`e}ncies de l'Espai (IEEC/CSIC), 
the Institut de F{\'i}sica d'Altes Energies, Lawrence Berkeley National Laboratory, the Ludwig-Maximilians Universit{\"a}t M{\"u}nchen and the associated Excellence Cluster Universe, 
the University of Michigan, the National Optical Astronomy Observatory, the University of Nottingham, The Ohio State University, the University of Pennsylvania, the University of Portsmouth, 
SLAC National Accelerator Laboratory, Stanford University, the University of Sussex, Texas A\&M University, and the OzDES Membership Consortium.

The DES data management system is supported by the National Science Foundation under Grant Number AST-1138766 and AST-1536171.
The DES participants from Spanish institutions are partially supported by MINECO under grants AYA2015-71825, ESP2015-88861, FPA2015-68048, SEV-2012-0234, SEV-2012-0249, and MDM-2015-0509, 
some of which include ERDF funds from the European Union. IFAE is partially funded by the CERCA program of the Generalitat de Catalunya.

We are grateful for the extraordinary contributions of our CTIO colleagues and the DECam Construction, Commissioning and Science Verification
teams in achieving the excellent instrument and telescope conditions that have made this work possible.  The success of this project also 
relies critically on the expertise and dedication of the DES Data Management group.


\appendix



\bibliographystyle{mnras}
\bibliography{database}

\section*{Affiliations}

$^{1}$ Kavli Institute for Particle Astrophysics \& Cosmology, P. O. Box 2450, Stanford University, Stanford, CA 94305, USA\\
$^{2}$ Institute of Space Sciences, IEEC-CSIC, Campus UAB, Carrer de Can Magrans, s/n,  08193 Barcelona, Spain\\
$^{3}$ Kavli Institute for Cosmological Physics, University of Chicago, Chicago, IL 60637, USA\\
$^{4}$ Institut de F\'{\i}sica d'Altes Energies (IFAE), The Barcelona Institute of Science and Technology, Campus UAB, 08193 Bellaterra (Barcelona) Spain\\
$^{5}$ Fermi National Accelerator Laboratory, P. O. Box 500, Batavia, IL 60510, USA\\
$^{6}$ Instituci\'o Catalana de Recerca i Estudis Avan\c{c}ats, E-08010 Barcelona, Spain\\
$^{7}$ SLAC National Accelerator Laboratory, Menlo Park, CA 94025, USA\\
$^{8}$ Department of Physics, University of Arizona, Tucson, AZ 85721, USA\\
$^{9}$ Center for Cosmology and Astro-Particle Physics, The Ohio State University, Columbus, OH 43210, USA\\
$^{10}$ Department of Physics, The Ohio State University, Columbus, OH 43210, USA\\
$^{11}$ Cerro Tololo Inter-American Observatory, National Optical Astronomy Observatory, Casilla 603, La Serena, Chile\\
$^{12}$ Department of Physics \& Astronomy, University College London, Gower Street, London, WC1E 6BT, UK\\
$^{13}$ Department of Physics and Electronics, Rhodes University, PO Box 94, Grahamstown, 6140, South Africa\\
$^{14}$ LSST, 933 North Cherry Avenue, Tucson, AZ 85721, USA\\
$^{15}$ CNRS, UMR 7095, Institut d'Astrophysique de Paris, F-75014, Paris, France\\
$^{16}$ Sorbonne Universit\'es, UPMC Univ Paris 06, UMR 7095, Institut d'Astrophysique de Paris, F-75014, Paris, France\\
$^{17}$ Laborat\'orio Interinstitucional de e-Astronomia - LIneA, Rua Gal. Jos\'e Cristino 77, Rio de Janeiro, RJ - 20921-400, Brazil\\
$^{18}$ Observat\'orio Nacional, Rua Gal. Jos\'e Cristino 77, Rio de Janeiro, RJ - 20921-400, Brazil\\
$^{19}$ Department of Astronomy, University of Illinois, 1002 W. Green Street, Urbana, IL 61801, USA\\
$^{20}$ National Center for Supercomputing Applications, 1205 West Clark St., Urbana, IL 61801, USA\\
$^{21}$ Department of Physics and Astronomy, University of Pennsylvania, Philadelphia, PA 19104, USA\\
$^{22}$ Department of Physics, IIT Hyderabad, Kandi, Telangana 502285, India\\
$^{23}$ Instituto de Fisica Teorica UAM/CSIC, Universidad Autonoma de Madrid, 28049 Madrid, Spain\\
$^{24}$ Department of Astronomy, University of Michigan, Ann Arbor, MI 48109, USA\\
$^{25}$ Department of Physics, University of Michigan, Ann Arbor, MI 48109, USA\\
$^{26}$ Institute of Astronomy, University of Cambridge, Madingley Road, Cambridge CB3 0HA, UK\\
$^{27}$ Kavli Institute for Cosmology, University of Cambridge, Madingley Road, Cambridge CB3 0HA, UK\\
$^{28}$ Universit\"ats-Sternwarte, Fakult\"at f\"ur Physik, Ludwig-Maximilians Universit\"at M\"unchen, Scheinerstr. 1, 81679 M\"unchen, Germany\\
$^{29}$ Astronomy Department, University of Washington, Box 351580, Seattle, WA 98195, USA\\
$^{30}$ Santa Cruz Institute for Particle Physics, Santa Cruz, CA 95064, USA\\
$^{31}$ Australian Astronomical Observatory, North Ryde, NSW 2113, Australia\\
$^{32}$ Argonne National Laboratory, 9700 South Cass Avenue, Lemont, IL 60439, USA\\
$^{33}$ Departamento de F\'isica Matem\'atica, Instituto de F\'isica, Universidade de S\~ao Paulo, CP 66318, S\~ao Paulo, SP, 05314-970, Brazil\\
$^{34}$ George P. and Cynthia Woods Mitchell Institute for Fundamental Physics and Astronomy, and Department of Physics and Astronomy, Texas A\&M University, College Station, TX 77843,  USA\\
$^{35}$ Department of Astronomy, The Ohio State University, Columbus, OH 43210, USA\\
$^{36}$ Department of Astrophysical Sciences, Princeton University, Peyton Hall, Princeton, NJ 08544, USA\\
$^{37}$ Jet Propulsion Laboratory, California Institute of Technology, 4800 Oak Grove Dr., Pasadena, CA 91109, USA\\
$^{38}$ Department of Physics and Astronomy, Pevensey Building, University of Sussex, Brighton, BN1 9QH, UK\\
$^{39}$ Centro de Investigaciones Energ\'eticas, Medioambientales y Tecnol\'ogicas (CIEMAT), Madrid, Spain\\
$^{40}$ School of Physics and Astronomy, University of Southampton,  Southampton, SO17 1BJ, UK\\
$^{41}$ Instituto de F\'isica Gleb Wataghin, Universidade Estadual de Campinas, 13083-859, Campinas, SP, Brazil\\
$^{42}$ Computer Science and Mathematics Division, Oak Ridge National Laboratory, Oak Ridge, TN 37831\\
$^{43}$ Institute of Cosmology \& Gravitation, University of Portsmouth, Portsmouth, PO1 3FX, UK\\
$^{44}$ Department of Physics, Stanford University, 382 Via Pueblo Mall, Stanford, CA 94305, USA\\


\bsp
\label{lastpage}
\end{document}